\documentclass[%
 aip,
 amsmath,amssymb,
 reprint,%
]{revtex4-1}

\usepackage{graphicx}% Include figure files
\usepackage{dcolumn}% Align table columns on decimal point
\usepackage{bm}% bold math
\usepackage[utf8]{inputenc}
\usepackage[T1]{fontenc}
\usepackage{mathptmx}

\begin{document}

\preprint{AIP/123-QED}

\title{Development and application of a $^3$He Neutron Spin Filter at J-PARC}
%\title{Development of a $^3$He Neutron Spin Filter at Japan Proton Accelerator Research Complex}
%\title{Development of a $^3$He Neutron Spin Filter at Material and Life Science Experimental Facility of J-PARC }
%\title{Development of a $^3$He Neutron Spin Filter at J-PARC }
% Force line breaks with \\
\def\affJAEA{J-PARC Center, Japan Atomic Energy Agency, 2-1 Shirane, Tokai 319-1195, Japan}
\def\affJAEANUC{Nuclear Science and Engineering Center, Japan Atomic Energy Agency, 2-1 Shirane, Tokai 319-1195, Japan}
\def\affKEK{Institute of Materials Structure Science, High Energy Accelerator Research Organization, 1-1 Oho, Tsukuba, Ibaraki 305-0801 Japan}
\def\affCROSS{Research Center for Neutron Science and Technology, Comprehensive Research Organization for Science and Society, Tokai, Ibaraki 319-1106, Japan}
\def\affIBADAI{Graduate School of Science and Engineering, Ibaraki University, 2-1-1 Bunkyo, Mito, Ibaraki 310-8512. Japan}
\def\affKUMATORI{Institute for Integrated Radiation and Nuclear Science, Kyoto University, Kumatori, Osaka 590-0494, Japan}
\def\affKMI{Kobayashi-Maskawa Institute, Nagoya University, Chikusa, Nagoya, Aichi 464-8601, Japan}

\def\affNAGOYA{Department of Physics, Nagoya University, Chikusa, Nagoya, Aichi 464-8601, Japan}

\author{T. Okudaira}
\affiliation{\affJAEA}
\author{T. Oku}
\affiliation{\affJAEA}
\affiliation{\affIBADAI}

\author{T. Ino}
\affiliation{\affKEK}

\author{H. Hayashida}
\author{H. Kira}
\affiliation{\affCROSS}
\author{K. Sakai}
\affiliation{\affJAEA}
\author{K. Hiroi}
\affiliation{\affJAEA}
\author{S. Takahashi}
\affiliation{\affIBADAI}
\affiliation{\affJAEA}

\author{K. Aizawa}
\affiliation{\affJAEA}

\author{H. Endo}
\affiliation{\affKEK}
\author{S. Endo}
\affiliation{\affJAEANUC}
\affiliation{\affNAGOYA}

\author{M. Hino}
\affiliation{\affKUMATORI}
\author{K. Hirota}
\affiliation{\affNAGOYA}
\author{T. Honda}
\affiliation{\affKEK}
\author{K. Ikeda}
\affiliation{\affKEK}
\author{K. Kakurai}
\affiliation{\affCROSS}
\author{W. Kambara}
\affiliation{\affJAEA}

\author{M. Kitaguchi}
\affiliation{\affNAGOYA}
\affiliation{\affKMI}

\author{T. Oda}
\affiliation{\affKUMATORI}

\author{H. Ohshita}
\affiliation{\affKEK}

\author{T. Otomo}
\affiliation{\affKEK}
\affiliation{\affIBADAI}

\author{H. M. Shimizu}
\affiliation{\affNAGOYA}

\author{T. Shinohara}
\affiliation{\affJAEA}
\author{J. Suzuki}
\affiliation{\affCROSS}

\author{T. Yamamoto}
\affiliation{\affNAGOYA}

\date{\today}% It is always \today, today,

\begin{abstract}
We are developing a neutron polarizer with polarized $^3$He gas, referred to as a $^3$He spin filter,  based on the Spin Exchange Optical Pumping (SEOP) for polarized neutron scattering experiments at Materials and Life Science Experimental Facility (MLF) of Japan Proton Accelerator Research Complex (J-PARC). A $^3$He gas-filling station was constructed at J-PARC, and several $^3$He cells with long spin relaxation times have been fabricated using the gas-filling station. A laboratory has been prepared in the MLF beam hall for polarizing $^3$He cells, and compact pumping systems with laser powers of 30~W and 110~W, which can be installed onto a neutron beamline, have been developed.  A $^3$He polarization of 85\% was achieved at a neutron beamline by using the pumping system with the 110~W laser. Recently, the first user experiment utilizing the $^3$He spin filter was conducted, and there have been several more since then. The development and utilization of $^3$He spin filters at MLF of J-PARC are reported.
\end{abstract}

\maketitle
\section{Introduction}
A $^3$He spin filter is a neutron polarization device in which polarized $^3$He gas is enclosed in a special glass cell that does not contain boron. $^3$He spin filters are widely used to polarize neutrons and analyze neutron spins in neutron scattering experiments because of its capabilities to polarize neutrons in broad energy range and cover wide solid angle~\cite{Babcock2016, Dhiman2017, Chen2017, Berna2007, Lee2016, Beecham2011}. Since $^3$He nuclei have a strongly spin-dependent neutron absorption cross section, neutrons passing through polarized $^3$He gas are polarized. $^3$He nuclei are polarized using laser techniques known as Spin Exchange Optical Pumping (SEOP)~\cite{SEOP} or Metastability Optical Pumping (MEOP)~\cite{MEOP}.  In SEOP, $^3$He gas, N$_2$ gas, and pure rubidium or rubidium and potassium are encapsulated within a glass cell. The Rb or Rb-K cells are heated to around 170$^\circ$C or 220$^\circ$C, respectively, to evaporate the alkali metals, and circularly polarized laser light irradiates the cell. Unpaired electrons of alkali atoms are polarized by optical pumping, and the spins of the electrons are exchanged with those of $^3$He nuclei by hyperfine interaction. A diode laser array with output power on the order of 100 W and a wavelength of 794.7~nm, which corresponds to the D$_1$ absorption line of Rb, is typically employed. The $^3$He polarization can be improved by using Rb-K compared to the pure Rb case because of the high optical pumping efficiency of Rb-K~\cite{Hybrid}. When the laser irradiation is stopped, the $^3$He polarization exponentially decays due to non-uniformity of the surrounding magnetic field, collisions of $^3$He atom with each other, and collisions of $^3$He atoms with impurities in the glass cell and a glass wall. 

{\it Ex-situ} and {\it {\it in-situ}} methods are used for $^3$He spin filters with SEOP. In the {\it ex-situ} method, a $^3$He cell is polarized in a pumping station away from a neutron beamline, transferred in a magnetic cavity to keep the $^3$He polarization, and installed to a neutron beamline.  Since the $^3$He polarization decays during a neutron experiment, the $^3$He cell needs to be periodically re-polarized and re-installed for long-running experiments. However, the advantage of using {\it ex-situ} method is that it can be used for a neutron beamline with limited space, and the installation of the instrument is straightforward. This is beneficial at spallation neutron sources, which generally have small experimental spaces. As there are many
experiments with short measuring time at intense neutron beam facilities, those may be done without re-polarizing the $^3$He cell. In order to keep high $^3$He polarization during the experiment, a long relaxation time of $^3$He polarization is required, which makes details of the fabrication process of the $^3$He cells important.

On the other hand, in the {\it in-situ} method, the $^3$He cell is installed along with a laser system onto a neutron beamline. This method is suitable for longer experiments because a stable $^3$He polarization can be maintained for weeks. However, a larger space compared to the space required for the {\it ex-situ} method is needed, and the safety regulations involving the use of high powered lasers on a neutron beamline must be considered. For details of the technique for optically polarized $^3$He and its application, please see a reviewed paper Ref~\cite{Gentile2017}.\\

At MLF of J-PARC, world class intense pulsed neutron beams are provided to 23 neutron beamlines, and many neutron scattering experiments are conducted~\cite{Nakajima2017}. In Japan, the fundamental development of $^3$He polarization technique based on SEOP began in the 1990s~\cite{Sato1994}, and further development is still ongoing by Ino et al. at High Energy Accelerator Research Organization~\cite{Ino2005, Ino2009, Ino2016, Ino2017, InoNMR}. An {\it in-situ} SEOP system dedicated for a polarized neutron spectrometer
POLANO~\cite{Nakajima2017, POLANO} at beamline No. 23 has been developed, and now under commissioning~\cite{Ino2016, Ino2017}. In order to promote user experiments using $^3$He spin filters on the other beamlines at MLF, we are carrying out the development of a $^3$He spin filters with both {\it in-situ} and {\it ex-situ} methods for their versatile uses on site at J-PARC. Experiments demonstrating their use have been performed on several beamlines. User experiments and fabrication of $^3$He cells have recently begun. In this paper, we report development and utilization of $^3$He spin filters at MLF of J-PARC.

\section{gas-filling station}
\begin{figure}[h]
\centering
\includegraphics[width=8cm,clip]{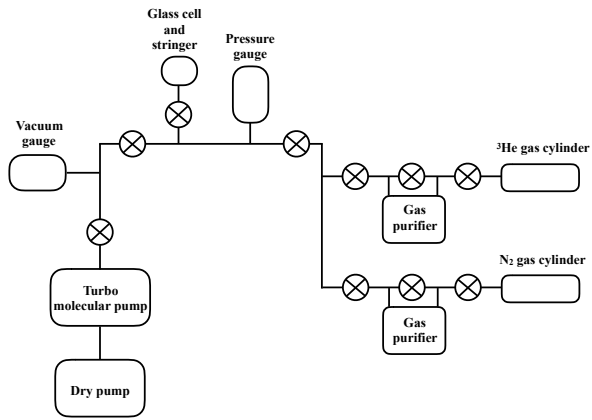}
\caption{The schematic diagram of the gas-filling station to fabricate $^3$He cells.}
\label{fig:gasstation}  
\end{figure}

\begin{figure}[h]
\centering
\includegraphics[width=8cm,clip]{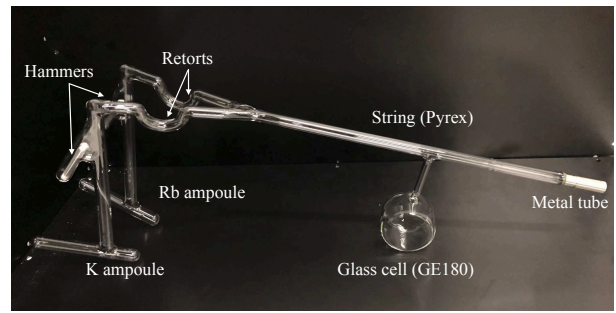}
\caption{The glassware used to fabricate a $^3$He cell. The ampoules are put in the parts labeled as K ampoule and Rb ampoule in the figure. The glass cell made of GE180 is attached to the string made of Pyrex at the top of the glass cell. The string is connected to the pumping station using a Swagelok connector at the metal tube. } 
\label{fig:string}  
\end{figure}

\begin{figure}[h]
\centering
\includegraphics[width=5cm,clip]{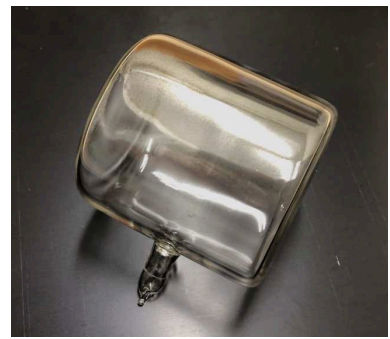}
\caption{$^3$He cell (Sekichiku) fabricated at J-PARC. } 
\label{fig:cell}  
\end{figure}

As mentioned in the previous section, a clean gas-filling station and fabrication process without impurities are important for a $^3$He cell with long relaxation time. The first gas-filling station was constructed in 2018 at J-PARC. The schematic diagram of the gas-filling station is shown in Fig.~\ref{fig:gasstation}. All the gas lines are wrapped with heaters for vacuum bake-out. An ultimate pressure of the gas-filling station was $5 \times 10^{-8}$~Pa after baking at 150$^\circ$C for a few days. \\
Boron-free aluminosilicate GE180 glass is widely used for the glass cell of a $^3$He spin filter due to a small wall relaxation effect and low permeability of $^3$He~\cite{Jianga2013, Chen2011,Berna2007, Salhi2014}. Our fabrication process of $^3$He cells is similar to those of NIST, JCNS, and ORNL~\cite{Jianga2013,Salhi2014, Chen2011}.  A cylindrical-shaped glass cell made of GE180 is attached to a glassware made of Pyrex glass, which is referred to as "string", (Fig.~\ref{fig:string}). The glass cell and the string are rinsed with neutral detergent, pure water, acetone, and alcohol before connecting to the gas-filling station. Rubidium and potassium ampoules are also rinsed with acetone and alcohol. After connecting the string to the gas-filling station, the ampoules are put in the string without breaking the ampoules, and the string is sealed off with a gas torch. The ampoules are broken in a nitrogen atmosphere by dropping hammers made of an iron rod covered with glass. The string and the glass cell are baked out for a week at 200$^\circ$C and 400$^\circ$C, respectively. In the next step, rubidium and potassium are distilled to retorts at 200$^\circ$C and 220$^\circ$C, respectively, by using the heaters. After that, the glass tube parts containing ampoules and hammers are pulled off with a gas torch. The glass cell and the string are baked out again for one week, and then the alkali metals in the retorts are distilled to the glass cell. Finally, $^3$He and N$_2$ gases purified using GC50 getters (SAES Getters) are filled to the glass cell, and the glass cell is sealed off by placing a gas torch at the Pyrex part. When encapsulating the gases above 1~atm, the glass cell is submerged in liquid nitrogen to keep the pressure inside the glass cell below 1~atm while sealing. So far, nine $^3$He cells have been fabricated with $^3$He pressures of 3.1~atm
using the gas-filling station at J-PARC, and four of them have found to have spin relaxation times over 150~hours. A list of these $^3$He cells and a photograph of Sekichiku cell are presented in Table~\ref{celllist} and Fig.~\ref{fig:cell}, respectively.

\begin{table}[h]
\caption{List of $^3$He cells using rubidium and potassium with long relaxation times at J-PARC. }
\begin{ruledtabular}
\begin{tabular}{cccc}
Name & Dimensions [mm]& $^3$He pressure [atm] & $T_1$ [h]\\
\hline
Chidori & $\phi$45 $\times$ 75 & 3.1 & 201\\
Karigane & $\phi$45 $\times$ 75 & 3.1 & 180\\
Sekichiku & $\phi$60 $\times$ 60 & 3.1 & 175\\
Hanabishi & $\phi$40 $\times$ 90 & 3.1 & 165\\
\end{tabular}
\end{ruledtabular}
\label{celllist}
\end{table}

\section{NMR and Pumping systems}
\subsection{$B_0$ coil and NMR system}
\begin{figure*}[htb]
\centering
\includegraphics[width=15cm,clip]{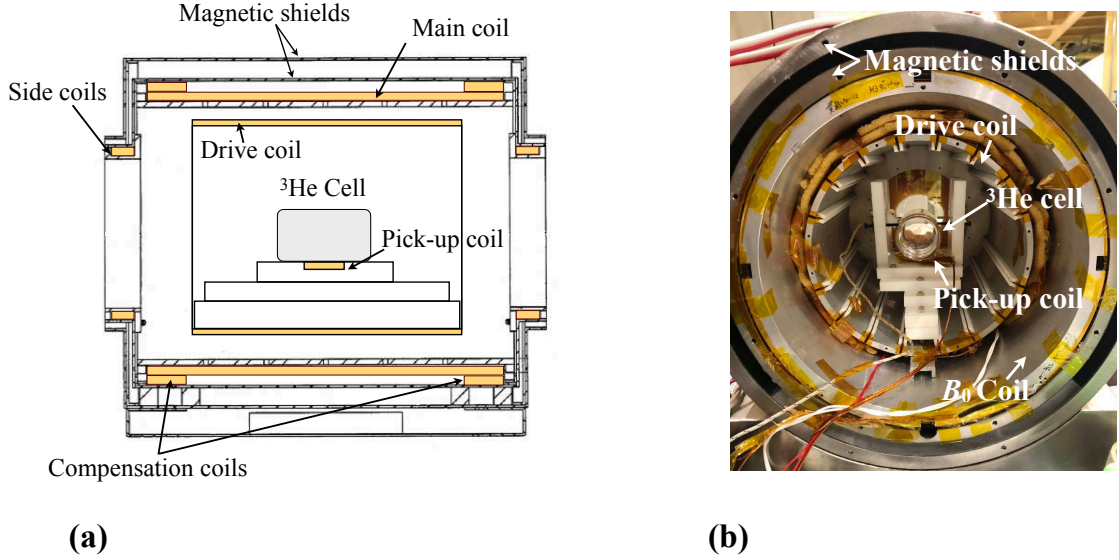}
\caption{(a) Cross-sectional view of the $B_0$ coil and the NMR coils. The magnetic shields are made from permalloy and its thicknesses are 2~mm. A static magnetic field is generated with the $B_0$ coil to keep $^3$He polarization. The drive coil is used to generate an oscillating magnetic field to flip $^3$He spins with AFP-NMR, and the spin flip signal is detected with the pick-up coil. (b) A photograph of the inside of the coil with the $^3$He cell and the NMR coils.} 
\label{fig:coils}  
\end{figure*}

 To keep the $^3$He polarization, a coil with double magnetic shields, referred to as $B_0$ coil, is mainly used for the laser pumping~\cite{Kira2013}. Figure~\ref{fig:coils} shows a cross-sectional view of the $B_0$ coil and a photograph of its interior.  The coil is consisting of a main solenoid, two compensation coils, and two side coils. The dimensions of the $B_0$ coil are 29.2~cm in diameter and 36.8~cm in length. 
 The $B_0$ coil was designed using the finite element method to make the field gradient less than 5$\times 10^{-4}$~/cm in a cylindrical region with a diameter of 10~cm and a length of 10~cm at the center of the coil. A magnetic field of 1.5~mT is produced to keep the $^3$He polarization. Frequency-sweep Adiabatic Fast Passage(AFP)-NMR was employed to flip the $^3$He spin as well as to evaluate the $^3$He polarization~\cite{InoNMR}. A cosine winding is used as a drive coil to apply an oscillating magnetic field to the $^3$He cell. The $^3$He precession signal is detected by a pick-up coil located under the $^3$He cell. The coils were installed inside the $B_0$ coil as shown in Fig.~\ref{fig:coils}. The polarization loss due to spin flip was measured as 3.8$\times$10$^{-5}$ per flip. The $^3$He cell is heated by electrical heating using a rubber heater with a power of 100~W wound around the drive coil during optical pumping.

\subsection{Pumping station at MLF}
\begin{figure*}[hbt]
\centering
\includegraphics[width=14cm,clip]{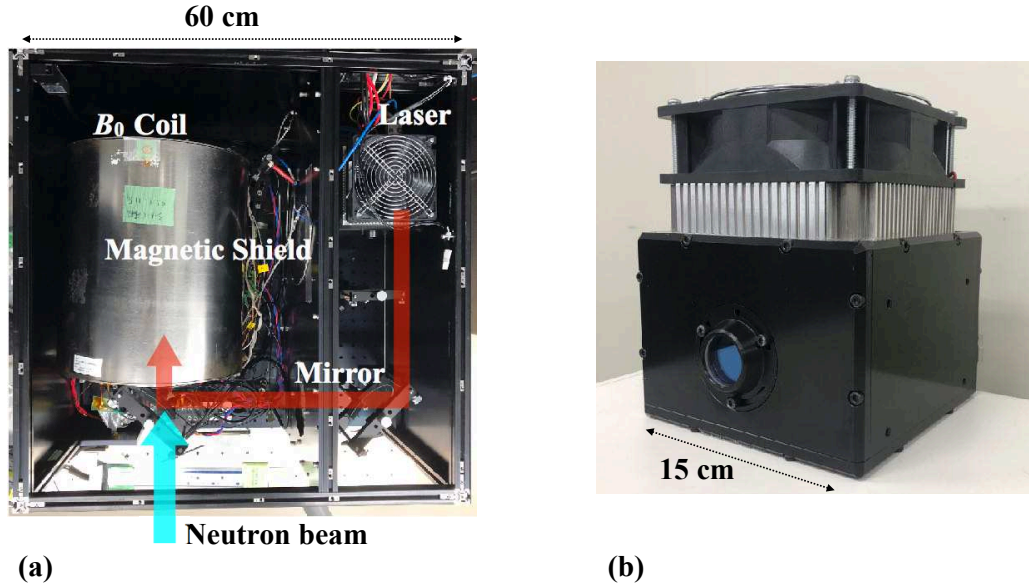}
\caption{(a) Compact pumping system which can be installed on a neutron beamline. The optical system has been constructed in the laser shield box. (b) Air cooling laser module. The laser power is 30~W.}
\label{fig:laser}  
\end{figure*}

We have set up a laboratory in the MLF beam hall to polarize $^3$He cells. $^3$He cells polarized at the laboratory can be provided to neutron beamlines in MLF. A pumping system using an air-cooling laser module with a laser power of 30\,W~\cite{Oku2015} has been installed in the laboratory (Fig.~\ref{fig:laser}). %It is placed in an area of the laboratory separated by a a laser safety curtain and is equipped with an interlock to turn off the laser when the laser shield curtain is opened, the lid of the laser shield box is opened, or the temperature of the $^3$He cell rises too high.
The dimensions of the pumping system are 60~cm $\times$ 60~cm $\times$ 40~cm, and it is used for both {\it in-situ} and {\it ex-situ} methods. 
Using a volume Bragg grating (VBG) element in the laser module, the center wavelength of the laser is tuned to 794.7~nm with a line width of 0.2~nm in Full Width at Half Maximum (FWHM). The laser light is circularly polarized using a quarter-wavelength plate installed in the module. The polarized laser light from the laser module is reflected by dielectric multi-layer coated quartz mirrors and irradiates the $^3$He cell. The temperature at the cell surface is kept about 170$^\circ$C for a pure Rb cell and about 210$^\circ$C for a cell with rubidium and potassium during optical pumping. A $^3$He polarization of $\sim$70\% is achieved with this laser system. A polarization measurement system using Electron Paramagnetic Resonance (EPR)~\cite{EPR} is also equipped. 

\subsection{Pumping system using a fiber laser}
\begin{figure*}[hbt]
\centering
\includegraphics[width=16cm,clip]{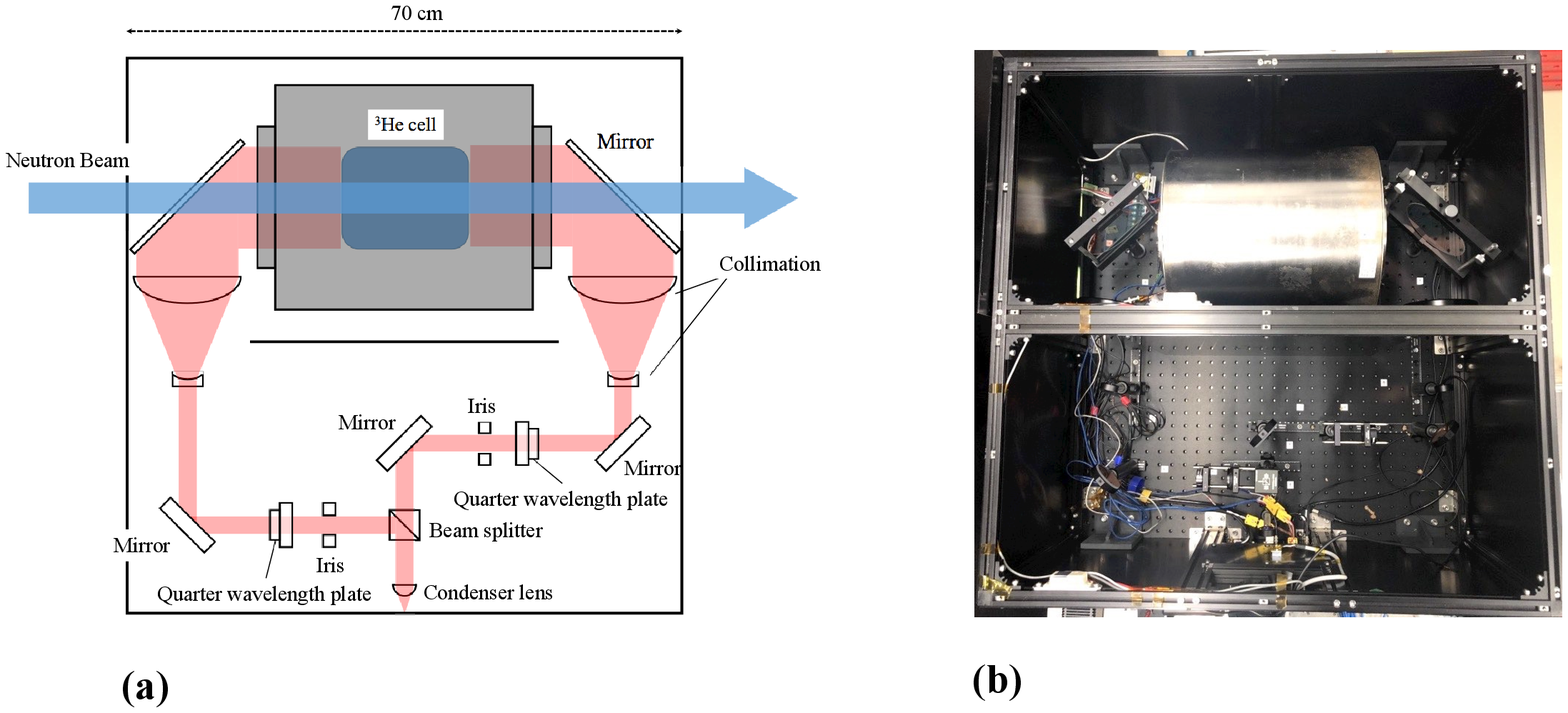}
\caption{(a) Schematic diagram of the pumping system using a fiber laser. (b) Photograph of the pumping system. The optical system was constructed in the laser shield box.}
\label{fig:Fiberlaser}  
\end{figure*}

In order to achieve higher $^3$He polarization, we have developed a separate pumping system using a fiber laser made by DILAS Diode Laser, Inc. (M1F4S22-794.7.[0,6]-100C-IS9), as shown in Fig.~\ref{fig:Fiberlaser}. The pumping system using this fiber laser has been constructed at another laboratory in the J-PARC research building, about 30\,m away from the MLF experimental hall. This fiber laser is an air-cooled diode laser with a maximum output power of 110~W. The center wavelength is tuned to 794.7~nm with a line width of 0.4~nm in FWHM using a VBG element. An optical fiber with an internal diameter of 400~$\rm{\mu}$m is connected to the laser shield box. An unpolarized laser beam from the fiber exit is split into two paths by a polarized beam splitter. The laser beams are circularly polarized with quarter-wavelength plates. The polarized laser beams irradiate both sides of a $^3$He cell. The degree of circular polarization in each path was measured using a polarimeter as more than 98\% at the $^3$He cell position. The dimensions of the pumping system are 70~cm $\times$ 70~cm $\times$ 40~cm, and it can also be used for the {\it in-situ} method. The pumping system is currently used only for the {\it ex-situ} method, and will be installed in the laboratory at MLF in the near future after following a safety review.

\section{Performance evaluation of the $^3$He spin filter}
\begin{figure}[h]
\centering
\includegraphics[width=8.5cm,clip]{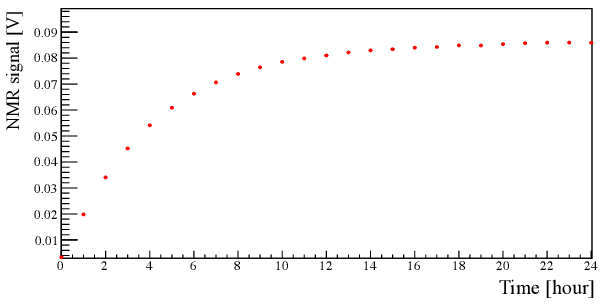}
\caption{Time evolution of the $^3$He polarization measured by AFP-NMR during the optical pumping. The $^3$He polarization saturates at about 24 hours after the start of the laser irradiation. The temperature at the surface of the cell was 205$^\circ$C during optical pumping.}
\label{fig:pumping}  
\end{figure}

We conducted an experiment at the NOBORU beamline~\cite{Nakajima2017, NOBORU}, which is a test neutron beamline for general purpose, to evaluate the $^3$He polarization for the Chidori cell, which has the longest $T_1$ as shown in Table~\ref{celllist}. The $^3$He cell was polarized using the pumping system with the fiber laser at J-PARC research building. Time evolution of the $^3$He polarization during optical pumping is shown in Fig.~\ref{fig:pumping}.

\begin{figure}[h]
\centering
\includegraphics[width=8cm,clip]{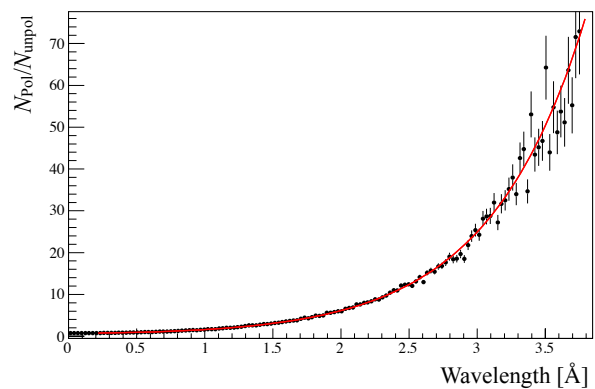}
\caption{Ratio of the numbers of transmitted neutrons for the polarized to the unpolarized $^3$He cell. The red line shows the best fit. The error bars show the statistical uncertainty.}
\label{fig:transmission}  
\end{figure}

\begin{figure}[h]
\centering
\includegraphics[width=8cm,clip]{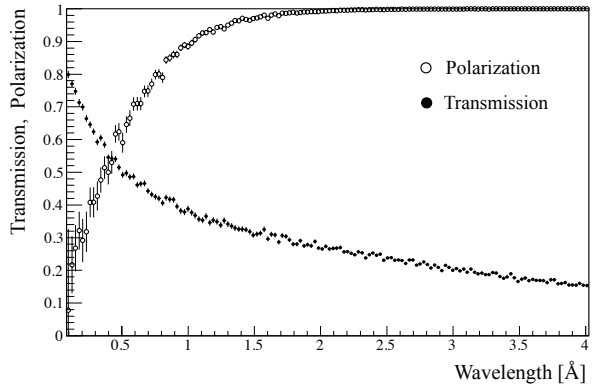}
\caption{Neutron wavelength dependences of neutron polarization (open circles) and neutron transmission (filled circles). }
\label{fig:PolTrans}  
\end{figure}

After the $^3$He polarization was saturated, the $^3$He cell, the $B_0$ coil, and the NMR system were transferred to the NOBORU beamline in MLF while applying a magnetic field to the $^3$He cell using a battery to keep the $^3$He polarization. Neutron transmission was measured using a Gas Electron Multiplier (GEM) detector~\cite{GEM} for the polarized and unpolarized $^3$He cell as well as for an empty glass cell with the same dimensions. The number of the transmitted neutrons through the $^3$He cell is described as~\cite{nPol}
\begin{eqnarray}
N_0T_{\rm{cell}}\epsilon(\lambda)\exp(-\rho d \sigma(\lambda))\cosh(P_{\rm{He}}\rho d \sigma(\lambda)),
\end{eqnarray}
where, $N_0$ is the number of incident neutrons, $T_{\rm{cell}}$ is the transmission of the cell windows, $\epsilon(\lambda)$ is the detection efficiency of the GEM detector, $P_{\rm{He}}$ is $^3$He polarization, $\rho$ is the number density of $^3$He nuclei, $d$ is the thickness of $^3$He gas, $\sigma(\lambda)$ is the absorption cross section of the $^3$He nucleus for unpolarized neutrons, and $\lambda$ is the wavelength of incident neutrons, which is calculated from the time of flight.
The ratio of transmitted neutrons for the polarized to unpolarized $^3$He cell $N_{\rm{pol}}/N_{\rm{unpol}}$ is described as   
\begin{eqnarray}
\frac{N_{\rm{pol}}}{N_{\rm{unpol}}}=\cosh(P_{\rm{He}}\rho d \sigma(\lambda)).
\label{eq:cosh}
\label{ratio}
\end{eqnarray}
The product of the $^3$He gas pressure and thickness $\rho d$ was obtained from the measurement of the ratio of transmitted neutrons for the unpolarized $^3$He cell to the empty glass cell as 20.8$\pm$0.1~atm$\cdot$cm. The measured ratios of $N_{\rm{pol}}/N_{\rm{unpol}}$ are plotted againt the neutron wavelength in Fig.~\ref{fig:transmission} along with the curve of best fit using Eq.~(\ref{eq:cosh}) with $P_{\rm{He}}$ as a free parameter.
The $^3$He polarization was measured to be 85.3$\pm$0.5\%. The uncertainty comes mostly from the measured $\rho d$. \\
The neutron wavelength dependences on the neutron polarization is obtained using the following equation: 
\begin{eqnarray}
P_n=\sqrt{1-\frac{T^2_0}{T^2}},
\label{eq:nPol}
\end{eqnarray}
where $P_n$ is the polarization of neutrons passed through the polarized $^3$He cell, $T$ is the transmission of polarized $^3$He cell,  and $T_0$ is the transmission of the unpolarized $^3$He cell. The transmissions $T_0$ and $T$ are described as 
\begin{eqnarray}
T_0=\frac{N_{\rm{unpol}}}{N_{\rm{cell}}}, \ \ T=\frac{N_{\rm{pol}}}{N_{\rm{cell}}},
\end{eqnarray}
where $N_{\rm{cell}}$ is the transmitted neutrons for the empty glass cell. The neutron wavelength dependence of the neutron polarization and the neutron transmission are shown in Fig.~\ref{fig:PolTrans}. The proton beam power to the neutron source was 500~kW, and each measurement took 15~min. The relaxation time $T_1$ on the beamline was 170~h, which was measured with AFP-NMR.

\section{Trial and User Experiments using $^3$He spin filter}
We have carried out several trial experiments with {\it in-situ} and {\it ex-situ} SEOP methods at neutron beamlines of MLF~\cite{Hayashida16, Hayashida19, Kira15, Sakai14}. In 2017, the $^3$He spin filter was first supplied to a user experiment. Although no user experiment was conducted in 2018, five user experiments were performed in 2019--2020. Neutron beamlines, where we conducted experiments using $^3$He spin filters, are listed in Table~\ref{beamlinelist}. In this section, we report on experiments performed on typical beamlines at MLF using $^3$He spin filters, including the results already reported.
\begin{table}[h]
\caption{List of neutron beamlines, where experiments were conducted with $^3$He spin filters.}
\begin{ruledtabular}
\begin{tabular}{ccccc}
No. &Name & Purpose of beamline& Type of $^3$He spin filter\\
\hline
04 & ANNRI & Nuclear reaction & {\it ex-situ} \\
05 & NOP & Fundamental physics & {\it ex-situ} \\
06 & VIN ROSE& Spin echo & {\it ex-situ} \\
10 & NOBORU & General purpose & {\it ex-}, {\it in-situ} \\
15 & TAIKAN & SANS & {\it ex-}, {\it in-situ} \\
17 & SHARAKU & Reflectometer  & {\it in-situ}  \\
18 & SENJU & Single crystal diffraction &{\it ex-situ}  \\
22 & RADEN & Pulsed neutron imaging &{\it in-situ}  \\
\end{tabular}
\end{ruledtabular}
\label{beamlinelist}
\end{table}
\subsection{BL10 NOBORU}
A trial experiment for magnetic imaging was carried out~\cite{Hayashida16} with {\it in-situ} and {\it ex-situ} systems using $^3$He cell with pure Rb and the 30~W laser as shown in Fig.~\ref{fig:laser}. The {\it in-situ} system with a $^3$He cell of 17~atm$\cdot$cm and the {\it ex-situ} system with $^3$He cell of 11~atm$\cdot$cm were used as a neutron spin polarizer and analyzer, respectively. The dimensions of the $^3$He cells were 35~mm in diameter and 55~mm in length for both. A coil with a 0.35 mm-thick ring-shaped magnetic steel core was placed between the polarizer and the analyzer as a sample. A two-dimensional RPMT neutron detector~\cite{Hirota2005} was set after the analyzer. The magnetic field applied to the sample was controlled with the current sent to the coil. Two-dimensional images of the neutron polarization were obtained, and obvious differences in the spin rotations of neutrons were observed as the wavelength dependence of the neutron polarization in two current conditions, $I$ = 1.42~A and $I$ = 0~A. Our result concluded that the magnetization patterns in these two conditions are different. \\\\
Furthermore, a study for magnetic atomic resolution holography with polarized neutrons using a $^3$He spin filter is currently underway. Two user experiments for this study were conducted at NOBORU in 2019--2020. 

\subsection{BL04 ANNRI}
\begin{figure}[h]
\centering
\includegraphics[width=8cm,clip]{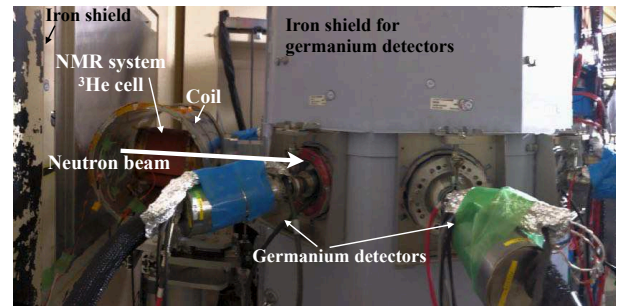}
\caption{Photograph of the {\it ex-situ} system installed at ANNRI beamline. The coil wrapped with permalloy sheets, NMR system, and $^3$He cell were placed between the iron shields.}
\label{fig:BL04}  
\end{figure}

BL04 ANNRI~\cite{Nakajima2017, ANNRI} is a neutron beamline used for studies of nuclear science such as nuclear data for nuclear technology, astrophysics, and quantitative analyses. A germanium detector assembly using 22 high-quality germanium crystals is installed on this beamline to measure the ($n$,$\gamma$) reaction. Measurements of the ($n$,$\gamma$) reaction for the study of fundamental symmetry violation in nuclei have been conducted as a user experiment using a $^3$He spin filter~\cite{Yamamoto2018,Yamamoto2019,Yamamoto20}. A $^3$He spin filter with the {\it ex-situ} system was used to polarize epithermal neutrons to measure the ($n$,$\gamma$) reaction with polarized neutrons at the 0.74~eV neutron resonance of $^{139}$La.  A lanthanum sample was placed at the center of the germanium detector assembly, and the emitted $\gamma$-rays from the sample were detected by the germanium detectors. The $^3$He cell, solenoid, and compensation coils with a magnetic shield made of permalloy were installed between the detector and a neutron collimator (Fig.~\ref{fig:BL04}). The magnetic field used to keep the $^3$He polarization was perpendicular to the neutron beam in order to obtain a vertically polarized neutron beam. The magnetic shield had two openings in the path of the neutron beam to avoid a decay of the neutron polarization. Neutron polarization was held by the magnetic filed produced with the guide magnets to a lanthanum target. A $^3$He cell using pure Rb with a pressure thickness of 19.3~atm$\cdot$cm and dimensions of 50~mm diameter and 70~mm length was used. The $^3$He cell was polarized with the 30\,W laser system at MLF, and the $^3$He polarization at the start of the measurement was typically 60\%. The relaxation time of the $^3$He polarization was 130~h. In 2019, as a result of three separate measurements using the $^3$He spin filter, a significant spin-dependent angular distribution of $\gamma$-rays from the 0.74~eV neutron resonance of $^{139}$La was observed~\cite{Yamamoto20}. \\
Furthermore, a ($n$,$\gamma$) reaction measurement using polarized epithermal neutrons to obtain nuclear data for reactor science is being planned, and the development of a $^3$He cell for epithermal neutrons is an ongoing process. 
\subsection{BL17 SHARAKU}
\begin{figure}[h]
\centering
\includegraphics[width=8cm,clip]{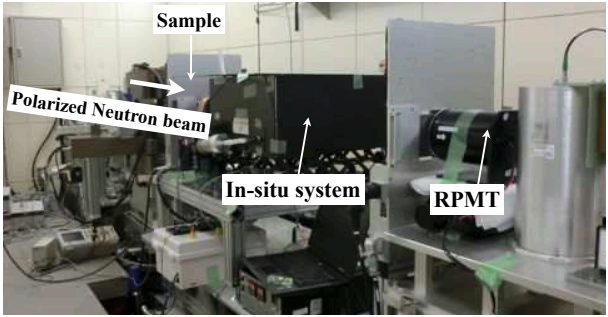}
\caption{Photograph of the {\it in-situ} system installed at SHARAKU beamline.}
\label{fig:BL17}  
\end{figure}

A trial experiment using the {\it in-situ} system on a beamline for neutron reflectometry: BL17 SHARAKU~\cite{Nakajima2017,SHARAKU} was performed~\cite{Hayashida16}. The SHARAKU beamline has a neutron polarizer and an analyzer consisting of supermirrors. Some of the supermirrors are stacked vertically in order to cover a wide area, which analyze the scattered neutron spins with wavelengths longer than 0.2 nm. However, some neutrons, which transmitted through the supermirrors are scattered by the supermirrors, and spacial distribution of neutrons reflected from a sample is disturbed. A $^3$He spin filter was used as the analyzer instead of the supermirrors to reduce uncertainty produced with the supermirrors (Fig.~\ref{fig:BL17}). The $^3$He cell using pure Rb with a pressure thickness of 11.1~atm$\cdot$cm, diameter of 35~mm, and length of 55~mm was used. A Fe-Cr multi-layered thin film was used as a sample, and off-specular measurements were performed using the {\it in-situ} system. The $^3$He polarization was 60\% during the experiment. The {\it in-situ} system successfully demonstrated and worked stably over 4 days of the experimental time. The sample had been measured previously at ISIS, and we obtained the identical results on SHARAKU. This satisfactory implies that our {\it in-situ} system worked as designed. 

\subsection{ BL06 VIN ROSE}
\begin{figure}[h]
\centering
\includegraphics[width=8cm,clip]{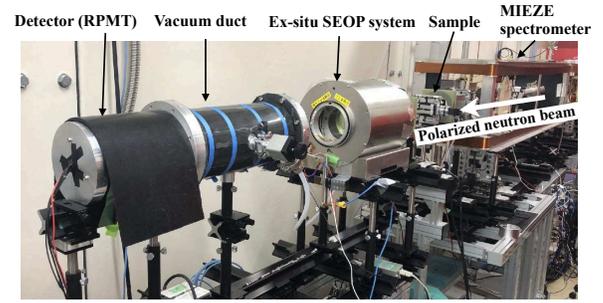}
\caption{Photograph of the {\it ex-situ} system installed on the VIN ROSE beamline. The polarized neutron beam was scattered by the sample, and the spin of the scattered neutron was analyzed with the $^3$He spin filter. The vacuum duct was wrapped with boron sheets to reduce background neutrons.}
\label{fig:BL06}  
\end{figure}

BL06 VIN ROSE~\cite{Nakajima2017, Hino2017} is equipped with a modulated intensity by zero effort (MIEZE) spectrometer to analyze the slow dynamics of condensed matter by measuring the intermediate scattering function. Experiments using a combination of a MIEZE spectrometer and the time of flight (TOF-MIEZE~\cite{VINROSE}) method have been carried out, and additionally, polarization analysis of scattered neutrons by a sample is planned in order to study spin dynamics by installing a spin analyzer. We installed a $^3$He cell after the sample position as an analyzer to confirm that the $^3$He spin filter can be used for the MIEZE spectrometer~\cite{Hayashida19}. The MIEZE signal was measured with and without the $^3$He cell, and no difference was observed in the signal contrast. This result implies that diffusion of $^3$He gas does not affect the MIEZE signal and a $^3$He spin filter can be used as a second analyzer in the MIEZE spectrometer. Furthermore, in 2020, an experiment to measure the spin dynamics of ferrofluid was conducted by using a $^3$He spin filter as shown in Fig.~\ref{fig:BL06}. The $^3$He polarization of 72\% and relaxation time of 120~h was achieved on this beamline in the experiment, and the analysis is currently underway.  
\subsection{BL15 TAIKAN}
\begin{figure}[h]
\centering
\includegraphics[width=8cm,clip]{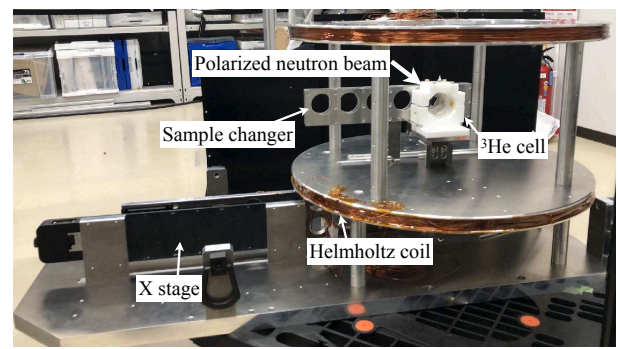}
\caption{Photograph of an {\it ex-situ} SEOP system for a SANS experiment at TAIKAN beamline. The {\it ex-situ} system is equipped with a sample changer which can be used to place up to six samples. The sample position is moved with the X-stage. The $^3$He polarization is kept by the Helmholtz coil. The $^3$He cell with the dimensions of $\phi$60$\times$40~mm is placed at 10~mm downstream of the sample to cover a large solid angle. }
\label{fig:BL15}  
\end{figure}

BL15 TAIKAN~\cite{Nakajima2017, TAIKAN} is a small and wide angle neutron scattering instrument. We conducted a trial experiment to measure coherent and incoherent scattering cross sections from hydrogen contained material by analyzing scattered neutrons using the {\it in-situ} system~\cite{Kira15}. Supermirrors were installed upstream of the beamline as a neutron polarizer, and the $^3$He spin filter was used as a spin analyzer. Silver behenate was used as the sample, and the $^3$He cell with pure Rb was placed 120~mm downstream of the sample. The diameter of $^3$He cell was 35~mm, the length was 55~mm, and the gas pressure was 3~atm. A stable $^3$He polarization of 68\% was kept during the experiment. Scattered neutrons reaching a part of the small-angle detector bank were spin analyzed. The coherent and incoherent scattering components were successfully separated by analyzing the scattered neutrons. 
Although the trial experiment was successful, the solid angle of the $^3$He cell was inadequate and the pressure-thickness was not suitable for the neutron wavelength used in TAIKAN. In order to cover a larger solid angle, an {\it ex-situ} SEOP system using a larger $^3$He cell, which is to be placed 10~mm downstream of the sample, is being prepared (Fig.~\ref{fig:BL15}). A new $^3$He cell with rubidium and potassium has been fabricated with a diameter of 60~mm, a thickness of 40~mm, and a gas pressure of 1.5~atm for cold neutrons. More trial and user experiments will be carried out in 2020. 

\subsection{BL21 NOVA}
\begin{figure}[h]
\centering
\includegraphics[width=8cm,clip]{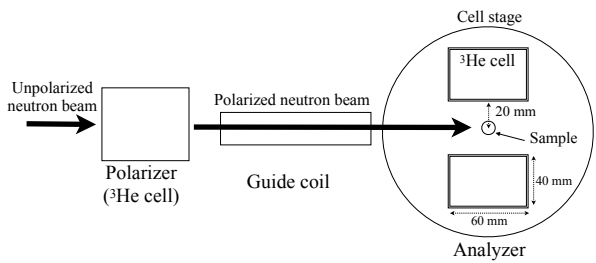}
\caption{Schematic view from the top of the {\it ex-situ} systems at NOVA. A neutron beam is polarized by the $^3$He cell installed upstream of the beamline, and the neutron polarization is held by the magnetic field produced with the guide coil to the sample. The scattered neutrons by the sample are analyzed with two $^3$He cells.}
\label{fig:BL21Config}  
\end{figure}

\begin{figure}[h]
\centering
\includegraphics[width=7cm,clip]{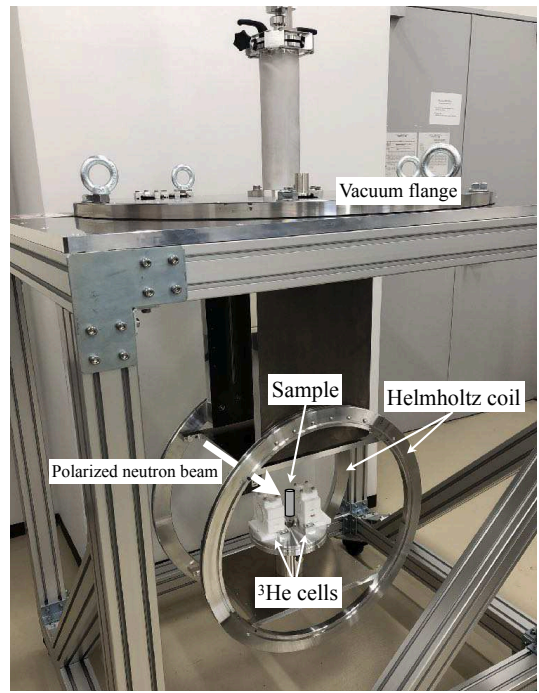}
\caption{Photograph of the {\it ex-situ} system for the analyzer. The sample, the $^3$He cells, and the Helmholtz coil are suspended from the vacuum flange. The $^3$He polarization is kept with the Helmholtz coil.}
\label{fig:BL21}  
\end{figure}

Experiments using $^3$He spin filters at BL21 NOVA~\cite{Nakajima2017, NOVA}, which is a total diffractometer, are also planned. NOVA is equipped with large solid angle neutron detectors. The detectors and a sample are installed in a large vacuum chamber. We are planning to install two {\it ex-situ} systems upstream of the beamline and around the sample to perform polarized neutron scattering experiments. The schematic view of the {\it ex-situ} systems at NOVA is shown in Fig.~\ref{fig:BL21Config}. The {\it ex-situ} system for the polarizer is installed at the upstream of the beamline, and the neutron polarization is held by the magnetic field produced with a guide coil to the sample. The scattered neutrons by the sample are analyzed with two $^3$He cells. This {\it ex-situ} system for the analyzer which can be installed to the vacuum chamber is developed as shown in Fig.~\ref{fig:BL21}. A sample is suspended from the top of the vacuum flange. The neutrons scattered by the sample are analyzed with two $^3$He cells placed at a distance of 20~mm from the sample. The stage for the $^3$He cells can be rotated so that the scattered neutrons of any angle can be analyzed. The $^3$He polarization is kept using the Helmholtz coil that produces 40~W of heat. The heat from the coil flows to the flange through the 30~mm thick aluminum plates connected to the coil, and the flange is air-cooled. A thermal simulation of the {\it ex-situ} system in a vacuum was performed. As a result of the simulation, the maximum temperature was determined to be 76$^\circ$C at the bottom of the coil, which does not cause a problem. A test experiment using the {\it ex-situ} system will be carried out in 2020. 

\section{Conclusion}
Development of $^3$He spin filters for user experiments are being carried out at MLF at J-PARC. The {\it in-situ} system using a 30~W laser has been developed, and several trial experiments using $^3$He cells with pure Rb have been conducted. Recently, a gas-filling station has been constructed at J-PARC and several high quality $^3$He cells using rubidium and potassium were fabricated. Additionally, the pumping system using a 110~W fiber laser was developed. High $^3$He polarization of 85\% was achieved on a neutron beamline using the $^3$He cell with rubidium and potassium and the pumping system using the fiber laser. The first user experiment utilizing a $^3$He spin filter took place in 2017, and five user experiments were conducted in 2019--2020. These experiments are beginning to yield scientific results. Preparations are underway for further user experiments using $^3$He spin filters.  

\section*{Acknowledgement}
The authors would like to thank the staff of each beamline, and MLF and J-PARC for operating the accelerators and the neutron production target. This work was supported by MEXT KAKENHI Grant Nos.~19K21047, JP19GS0210, JSPS KAKENHI Grant Nos.~18K11929, JP18H05518, and JP17H02889. The neutron scattering experiments at BL06 and BL21 were approved by the Neutron Scattering Program Advisory Committee of IMSS, KEK (Proposal Nos. 2019S06 and 2019S07). The neutron scattering experiments at BL04 were performed under the user program of MLF (Proposals Nos.~2018B0148 and 2019A0185). The neutron scattering experiments at BL10, BL15, and BL17 were performed under the program of project use (Proposal Nos.~2012P0802, 2014P0802, and 2019P0201).
\bibliography{He3}% Produces the bibliography via BibTeX.
\end{document}